\RequirePackage{fix-cm}
\documentclass[onecolumn,epjc3]{svjour3}
\smartqed  
\RequirePackage{graphicx}
\usepackage[caption=false]{subfig}
\usepackage{subfloat}
\usepackage{slashed}
\usepackage{amsmath,mathtools}
\usepackage{setspace}
\usepackage{xcolor}
\begin{document}

\title{Quasibound states for a scalar field under the influence of an external magnetic field in the near-horizon geometry of the BTZ black hole with torsion}

\author{ \resizebox{7.5cm}{!}{$ \textbf{Abdullah\ Guvendi}\thanksref{e1,a1},\quad  \textbf{Semra\ Gurtas\ Dogan}\thanksref{e2,a2}$}}


\thankstext{e1}{e-mail: abdullah.guvendi@erzurum.edu.tr; abdullahguvendi@gmail.com}
\thankstext{e2}{e-mail: semragurtasdogan@hakkari.edu.tr (Corresponding author)}


\institute{Department of Basic Sciences, Faculty of Science, Erzurum Technical University, 25050, Erzurum, Turkey \label{a1} \and Department of Medical Imaging Techniques, Hakkari University, 30000, Hakkari, Turkey \label{a2}}

\maketitle
\begin{spacing}{2}
\begin{abstract}
We consider a charged scalar field under the effect of an external uniform magnetic field in the near-horizon geometry of the Banados-Teitelboim-Zanelli black hole with torsion and obtain quasi-stationary states of the system under consideration through obtaining analytical solution of the corresponding Klein-Gordon equation. We obtain the solution function of the equation and accordingly we arrive at a complex spectra. We observe that the real oscillation frequency of the modes and their decay time depends on the strength of the external magnetic field besides the parameters of the geometric background. We see that the amplitude of the real oscillation modes decreases and the decay time of the modes becomes longer as the strength of the external magnetic field increases. The results also indicate that the geometric background is stable under such a perturbation field.

\end{abstract}

\section{Introduction}\label{sec:1}

General relativity in 2+1 dimensions has gained great interest after the seminal papers of Deser, Jackiw, 't Hooft \cite{deser1984three,deser1988classical,t1988non} and Witten \cite{witten19882+,witten1989topology} were announced \cite{carlip19952+}. After these researches, the $2+1$ dimensional gravity has been regarded as a very useful laboratory to discuss some conceptual issues. At that times, it was believed that 2+1 dimensional gravity models cannot give much insight into $3+1$ dimensional real gravitating physical systems \cite{carlip19952+} due to the fact that three dimensional general relativity has no propagating degrees of freedom and has no Newtonian limit \cite{carlip19952+,barrow1986three}. However, it has been shown that the solution of the standard Einstein equations in asymptotically anti-de Sitter spacetime in 2+1 dimensions \cite{banados1992black,konoplya2011quasinormal,cvetkovic2019near} provides a black hole solution, the Banados-Teitelboim-Zanelli (BTZ) black hole (BH). This 2+1 dimensional BH is characterized through the mass, angular momentum and charge and is similar to its $3+1$ dimensional counterparts \cite{banados1992black} because of the fact that it has an event and an inner horizon (in the rotating case). However, the BTZ BH differs from the Kerr and Schwarzschild BHs in some aspects. For example, it has no curvature singularity at the spatial origin and is not asymptotically flat \cite{carlip19952+,banados1992black}. Because it is relatively simple, the BTZ BH has attracted great interest from many authors \cite{rincon2018scale,gecim2018quantum,xu2020diagnosis,canate2020nonlinear,panotopoulos2017greybody,tekincay2021exotic,panotopoulos2018quasinormal,suggestion,fathi2021adiabatic} and moreover has played a crucial role in many developments in string theory \cite{sfetsos1998microscopic,strominger1998black,birmingham2001exact}. The BHs are generic predictions of Einstein's general relativity but it was shown that these objects are not just mathematical objects. Hawking has shown us that the BHs may emit radiation from the horizon \cite{hawking1974black,hawking1975particle}. After these seminal works, the BHs have attracted great attention due to the fact that they have come to be regarded as an excellent laboratory for testing the theories of gravity. One of the useful ways to test these theories is to determine the characteristic oscillation modes of the BH spacetimes through using the techniques of Quantum Field Theory in curved spaces. It was shown that a BH spacetime spacetime undergoes damped oscillations when they are perturbed by scalar or vector fields \cite{regge1957stability,zerilli1970effective,zerilli1974perturbation,Chandrasekhar}. Determining quasinormal modes (QNMs) \cite{konoplya2011quasinormal} for the BHs, described by some parameters such as the mass, charge and angular momentum, is very important since these modes carry information about the BHs \cite{panotopoulos2018quasinormal,suggestion}. The QNMs can include long-lived trapped modes called as quasibound states (QBSs). These states, known also as quasi-stationary levels or resonance spectra, are localized in the BH potential well and tend to zero at spatial infinity. This means that there exist two boundary conditions associated with the QBSs \cite{berti2009quasinormal}. QBSs for many BHs have been widely studied and we need to mention a few investigations announced previously in the literature. The quasispectrum of resonant frequencies in the Schwarzschild acoustic BH spacetime \cite{Vieira}, QBSs of a massive scalar field for dilatonic charged BHs \cite{Yang}, resonance modes for the near-extremal Kerr BH spacetime \cite{Hod}, QBS test field configurations of scalar fields in the background of the Reissner-Nordström BH \cite{Marco} can be considered among such investigations. Also, one of the fundamental results of Kerr/CFT correspondence is that near-horizon geometry of the BHs encodes many important pieces of information \cite{guica2009kerr}. This means that determining the evolution of scalar or vector fields in the near-horizon geometry of the BHs with and without torsion can be very important. Here, it may be also useful to give some information about the role of torsion in the gravity theories \cite{cvetkovic2019near}. In Einstein's general relativity there is no torsion in the geometric backgrounds generated by source(s). However, it has been shown that there can be several new possibilities when the presence of the torsion or vorticity in the gravity theories are considered (see also Refs. \cite{cunha2016relativistic,vitoria2018rotating,zare2020non,guvendi2021relativisticb,guvendi2021effects}). For example, it seems to be possible that dark matter may represent a manifestation of non-trivial torsion \cite{cvetkovic2019near,shie2008torsion,belyaev2017torsion,tilquin2011torsion}, or dark energy and dark matter may be replaced by non-trivial torsion \cite{cvetkovic2019near,minkevich2009accelerating}.

On the other hand, it is known that magnetic fields exist at almost every point in the universe and these fields can magnetize the universe \cite{minkevich2009accelerating,enqvist1994ferromagnetic,grasso2001magnetic}. The origin of intra-cluster, galactic and cosmological magnetic fields is not yet exactly known. However, it is thought that dynamo effects in turbulent fluids can amplify exponentially the seed fields \cite{minkevich2009accelerating,enqvist1994ferromagnetic,grasso2001magnetic}. The magnetic fields can be responsible for many interesting phenomena in the universe and determining the effects of magnetic fields on the quantum mechanical systems has attracted great interest from authors in many areas of modern physics \cite{guvendi2021relativisticC,guvendi2021relativistic,dietl2008new,furtado1994landau,medeiros2012relativistic,hammad2020landau}. At high energies, determining the dynamics of quantum mechanical systems in curved spaces can be acquired by solving the Lorentz-invariant wave equations, such as the Dirac equation, Duffin-Kemmer-Petiau equation, vector boson equation, fully-covariant many-body equations and Klein-Gordon equation $(\mathcal{KG}E)$. Several applications of these equations can be found in the Refs. \cite{sakalli2004solution,hassanabadi2017duffin,guvendi2021vector,guvendi2020interacting,ahmed2020generalized,dogan2021two,guvendi2021Dynamics,guvendinoninertial}. Exact solutions of the relativistic wave equations are of high importance for many areas of modern physics. The $\mathcal{KG}E$ is used to describe the relativistic dynamics of spinless particles \cite{santos2018relativistic} and has been studied many times to analyze the characteristic oscillations of the geometry of the BHs \cite{cardoso2001scalar,horowitz2000quasinormal,konoplya2006stability,konoplya2002decay,detweiler1980klein} (for more details see the reviews in \cite{konoplya2011quasinormal,berti2009quasinormal}). We think that it can be useful to determine the QBSs for a charged scalar field under the influence of an external magnetic field in the near-horizon geometry of the BTZ BH with torsion by obtaining an analytical solution of the corresponding $\mathcal{KG}E$ (see also \cite{KGO}). The QBSs are obtained through solving the wave equations for ongoing waves at the exterior event horizon. These modes tend to zero at spatial infinity and the resonant frequency spectrum is related to the decay of the perturbation. That is, they are related to the damped oscillations. Such an investigation may allow us to discuss the effect of an external magnetic field on both the real oscillation frequency of the modes and their decay time. Here, we will study the dynamics of a test scalar field (massive and charged) exposed to an external uniform magnetic field in the near-horizon geometry of the BTZ BH with torsion (see \cite{cvetkovic2019near}). This manuscript is structured as follows: in  sec. \ref{sec:2} we write the generalized $\mathcal{KG}E$ and obtain the corresponding wave equation for a massive $\mathcal{KG}$ particle (with charge $e$) exposed to an external uniform magnetic field in the near-horizon geometry of the BTZ BH with torsion. In  sec. \ref{sec:3}we obtain a complex spectra for the test field in question by analytically solving the associated wave equation. In  sec. \ref{sec:4} we give a summary and discuss the results. In this manuscript, we use the units $\hbar=c=1$.

\section{Mathematical procedure}\label{sec:2}

In this part of the paper, we introduce the generalized $\mathcal{KG}E$ and obtain the corresponding form of this equation for a  spinless particle (massive and charged) exposed to an external magnetic field in the near-horizon geometry of the BTZ BH with torsion. The generalized form of the $\mathcal{KG}E$ can be written as the following \cite{zare2022interaction,turimov2019quasinormal,konoplya2008quasinormal,wu2015decay,kokkotas2011quasinormal}
\begin{eqnarray}
&\left[ \frac{1}{\sqrt{|-g|}}\mathcal{D}_{\lambda}\left( \sqrt{|-g|} g^{\lambda \nu} \mathcal{D}_{\nu}\right) -  \mu^{2}\right]\chi\left(\textbf{x} \right)=0,\nonumber\\
&\mathcal{D}_{\lambda}=\partial_\lambda +ie\mathcal{A}_\lambda,\quad \mathcal{D}_{\nu}=\partial_\nu +ie\mathcal{A}_\nu,\quad \lambda,\nu=0,1,2. \label{eq1}
\end{eqnarray}
Here, $g=det(g_{\lambda \nu})$, $g^{\lambda \nu} $ is the contravariant metric tensor, $e$ is the electrical charge of the particle, $\chi$ is the $\mathcal{KG}$ field with mass of $\mu$, $\mathcal{A}_\lambda$ is the electromagnetic vector potential and $\textbf{x}$ is the spacetime position vector of the particle. The near-horizon geometry of the rotating BTZ BH with torsion was investigated to analyze the role of torsion in the near-horizon geometry and it was shown that this spacetime background corresponds to the generalization of AdS self-dual orbifold possessing non-trivial torsion. More detailed discussion about this spacetime structure can be found in the Ref. \cite{cvetkovic2019near}. The mentioned geometric background can be represented through the following $3D$ line element with signature $+,-,-$ \cite{cvetkovic2019near}
\begin{eqnarray}
ds^2=\frac{4r_{0}r}{ \ell}dtd\phi-\frac{ \ell ^2}{4r^2}dr^2-r_{0}^2d\phi^2, \label{eq2}
\end{eqnarray}
for which $det(g_{\lambda \nu})=r_{0}^2$. In this metric, $r_{0}$ stands for the value of the radius of the event horizon and the torsion in the geometric background has been associated with $\ell$ (inverse) \cite{cvetkovic2019near}. According to the metric in Eq. (\ref{eq2}), the contravariant metric tensor is determined as the following
\begin{eqnarray}
g^{\lambda\nu}=\left(
\begin{array}{ccc}
\frac{\ell^2}{4r^2}& 0 & \ \ \ \frac{\ell}{2r_{0}r}\\
0&  -\frac{4r^2}{\ell^2} &\ \ \ 0 \\
\frac{\ell}{2r_{0}r}& 0 &\ \ \ 0
\end{array}%
\right).\label{eq3}
\end{eqnarray}
Here, it is useful to underline that we consider a charged scalar field under the effect of an external uniform magnetic field in the background geometry represented through the metric in the Eq. (\ref{eq2}) without discussing the origin of the external field. This field is taken into account through the angular component of the electromagnetic vector potential, as $\mathcal{A}_{\phi}=\frac{\mathcal{B}_{0}r}{2}$  \cite{dogan2019quasinormal}, in which $\mathcal{B}_{0}$ is the amplitude of the external uniform magnetic field and $r$ is the radial coordinate of the particle in question. The line element in Eq. (\ref{eq2}) allows to the factorization of the wave function, $\chi(\textbf{x})$, as $\chi(\textbf{x})=e^{-i\left(\omega t-m \phi\right)}\tilde \chi(r)$, in which $\omega$ and $m$ are the relativistic frequency and the azimuthal quantum number of the particle in question. Accordingly, one can obtain the following wave equation
\begin{eqnarray}
\left[\frac{d^2}{dr^2} + \frac{2}{r}\frac{d}{dr} +\frac{\mathcal{K}-\omega\mathcal{K}_1}{r^2}-\frac{\omega\mathcal{K}_2}{r^3}+\frac{\omega^2\mathcal{K}_3}{r^4}\right]\tilde \chi(r)=0,\label{eq4}
\end{eqnarray}
in which
\begin{eqnarray*}
\mathcal{K}=\frac{\mu^2 \ell^2}{4},\quad\mathcal{K}_1=\frac{\mathcal{B}\ell^3}{4r_0},\quad \mathcal{K}_2=\frac{m\ell^3}{4r_0},\quad \mathcal{K}_3=\frac{\ell^4}{16},
\end{eqnarray*}
and $\mathcal{B}=\frac{e\mathcal{B}_0}{2}$. In the next step, we will solve this equation and will obtain a frequency spectra for the considered perturbation field.

\begin{figure}[!h]
\centering
\includegraphics[width=1\textwidth]{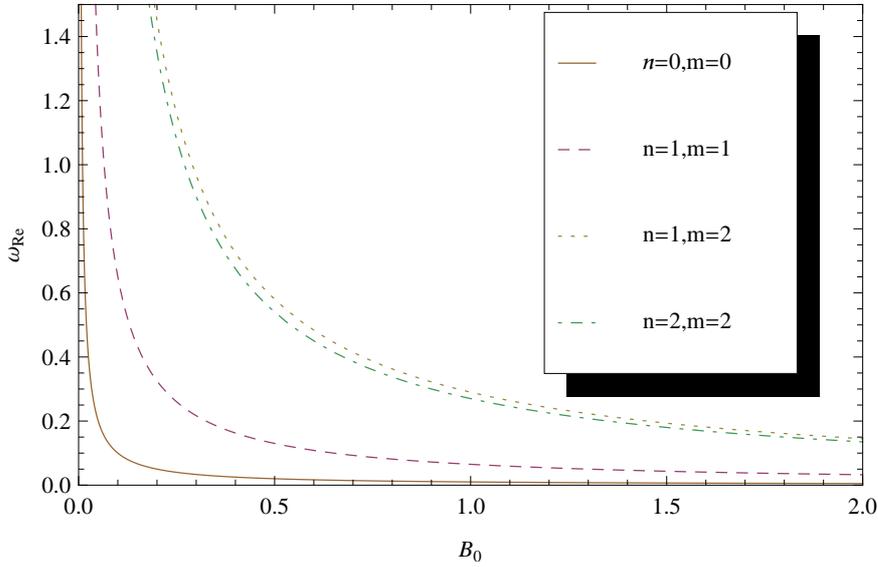}
\caption{ Dependence of the real oscillation frequencies ($\omega_{Re}$) on the strength of the external magnetic field. Here, $|e|=1,\mu=1$, $\ell=10$ and $r_0=10$.. }\label{fig1}
\end{figure}

\begin{figure}[!h]
\centering
\includegraphics[width=1\textwidth]{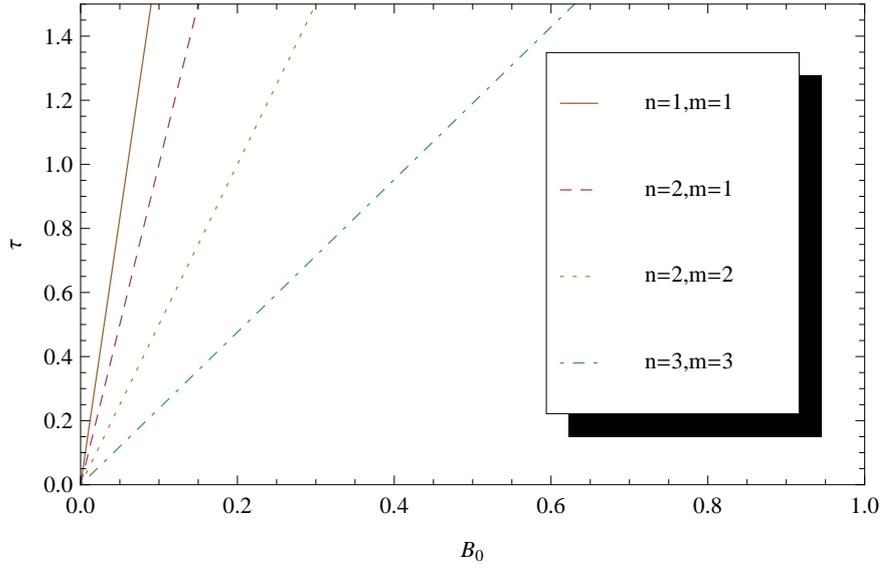}
\caption{Dependence of the decay times on the strength of the external magnetic field. Here, $|e|=1,\mu=1$, $\ell=10$ and $r_0=10$.}\label{fig3}
\end{figure}

\section{Quasibound states}\label{sec:3}

By considering a new change of variable, $\rho=\frac{2i\omega\sqrt{\mathcal{K}_3}}{r}$ (note that $\rho\rightarrow 0$ when $r\rightarrow \infty$), the wave equation in Eq. (\ref{eq4}) can be clarified as the following
\begin{eqnarray}
\left[\frac{d^2}{d \rho^2}+\frac{i\mathcal{K}_2}{2\sqrt{\mathcal{K}_3}\rho}+\frac{\mathcal{K}-\omega\mathcal{K}_1}{\rho^2}-\frac{1}{4}\right]\chi(\rho)=0. \label{eq5}
\end{eqnarray}
This equation has an irregular singular point at spatial infinity and regular singular point at spatial origin \cite{book1}.
Here, it is clear that information about the frequency of the system in question is lost when $\mathcal{B}_{0}=0$. For $\mathcal{B}_{0}=0$ case, it is not seem to be possible to determine the characteristic oscillations of the geometric background under such a test scalar field perturbation. This means also that the presence of the external uniform magnetic field allows to determine the frequency modes for the system under scrutiny. In case of $\mathcal{B}_{0}\neq0$, solution functions of this equation are obtained as Whittaker functions and they can be expressed in terms of the Confluent Hypergeometric ($\mathcal{CH}$) functions, ${}_1\mathcal{F}_{1}$. Thereby, the solution functions can be written as follows \cite{book1}

\begin{eqnarray}
\begin{split}
&\chi(\rho)= e^{-\frac{\rho}{2}}\rho^{-\frac{1}{2}} \left[\mathcal{C}_{1} \rho^{\beta}{}_1\mathcal{F}_{1}\left(\frac{1}{2}-\alpha+\beta, 1+2\beta, \rho\right)\right.\\
&+\left.\mathcal{C}_{2} \rho^{-\beta}{}_1\mathcal{F}_{1}\left(\frac{1}{2}-\alpha-\beta, 1+2\beta, \rho\right)\right],  \label{eq6}
\end{split}
\end{eqnarray}
\begin{eqnarray*}
\alpha=\frac{i\mathcal{K}_2}{2\sqrt{\mathcal{K}_3}},\quad \beta= \frac{\sqrt{1+4\left(\omega\mathcal{K}_1-\mathcal{K}\right)}}{2},
\end{eqnarray*}
where $\mathcal{C}_{1}$ and $\mathcal{C}_{2}$ are the normalization constants. In the near horizon, solution functions in the Eq. (\ref{eq6}) become as
\begin{eqnarray}
\chi(\rho)=\mathcal{C}_{1} \rho^{\beta-\frac{1}{2}}+\mathcal{C}_{2} \rho^{-\beta-\frac{1}{2}}. \label{eq6b}
\end{eqnarray}
Of course, near horizon solution ($\rho \rightarrow 0$) must obey the regularity. Hence, we get that $\mathcal{C}_{2}=0$. By using the asymptotic behaviors of the $\mathcal{CH}$ functions \cite{book1}, the solution function can be written (for $ \rho\rightarrow  \infty)$ as the following
\begin{eqnarray}
\begin{split}
&\chi(\rho) \rightarrow  \mathcal{C}_{1}\left(e^{\frac{1}{2}\rho}\rho^{-1-\alpha} \frac{\Gamma(1+2\beta)}{\Gamma(\frac{1}{2}+\beta-\alpha)}+ e^{-\frac{1}{2}\rho}\rho^{-1+\alpha}(-1)^{-\frac{1}{2}+\beta+\alpha}  \right). \label{eq6c}
\end{split}
\end{eqnarray}
The first term in the Eq. (\ref{eq6c}) corresponds to asymptotic ingoing waves, whereas the second term corresponds to asymptotic outgoing waves. The resonance condition, which requires that the following condition is satisfied $\frac{1}{2}+\beta-\alpha=-n$ \cite{book1} where $n=0,1,2..$ is the overtone number, guarantees that the solution function becomes polynomial of degree $n$ with respect to the variable $\rho$. This termination gives the following spectra \cite{guvendi2021relativistic}
\begin{eqnarray}
\omega=\omega_{Re}-i\omega_{Im},\label{eq7}
\end{eqnarray}
where
\begin{eqnarray*}
\omega_{Re}&=&\frac{\mu^2 m^2r_0-4r_0}{2e\mathcal{B}_0 \ell^3}+\frac{2r_0 n^{*^2}}{e\mathcal{B}_0 \ell^3}-\frac{8m^2}{e\mathcal{B}_0 r_0 \ell},
\end{eqnarray*}
and
\begin{eqnarray*}
\omega_{Im}&=&\frac{4mn^{*}}{e\mathcal{B}_0 \ell^2},\quad n^{*}=(n+\frac{1}{2}),
\end{eqnarray*}
which tell us what the dependence of the modes on both the strength of the external magnetic field and spacetime parameters. It can be seen that the spectrum consists of real ($\omega_{Re}$) and imaginary ($\omega_{Im}$) parts. Here, it is very important to say that our aim is to find discrete frequency modes of the system. Also, it is worth mentioning that we consider a charged scalar particle exposed to an external uniform magnetic field in the near-horizon geometry of the BTZ black hole with torsion. The line element in the Eq. (\ref{eq2}) describes only the near-horizon geometry of the BTZ black hole with torsion and hence the frequency expression given by Eq. (\ref{eq7}) cannot give all of the information about the extremal BTZ black hole. Our results describe only the interaction of a charged scalar particle exposed to an external uniform magnetic field with the near-horizon geometry of the BTZ black hole with torsion. However, we think that the near-horizon geometry of the black holes encodes important info about the black hole. Therefore our results may give some useful information about the BTZ black hole with torsion. The presence of a non-vanishing inward current at the BH horizon implies that these state cannot be steady states. These states, called as quasibound states, include real and imaginary parts. The imaginary part gives the exponential behavior, that is the field decays for $\omega_{Im}>0$ and grows for $\omega_{Im}<0$ in time. The $\omega_{Im}<0$ corresponds to unstable mode, whereas $\omega_{Im}> 0$ corresponds to the stable mode. Physically, positive $\omega_{Im}$ implies that such a charged particle has probabilities to permeate into the BH for non-vanishing strength of the external magnetic field. Here, we see that the real and imaginary parts of the spectra are inversely proportional to the magnetic field. As we mentioned before, our results are valid only for non-vanishing values of the strength of the external magnetic field (see also Refs. \cite{turimov2019quasinormal,konoplya2008quasinormal,wu2015decay,kokkotas2011quasinormal}). The real part $\omega_{Re}$ gives the real oscillation frequency of the modes and accordingly the imaginary part $\omega_{Im}$ relates with their damping rates \cite{konoplya2011quasinormal}. In Eq. (\ref{eq7}), we observe that real oscillation frequency of the modes and their decay times \cite{konoplya2011quasinormal} ($\tau$, $\tau\propto \frac{1}{\omega_{Im}}$) depend on the strength of the external uniform magnetic field besides the parameters of the geometric background, $r_{0}$ and $\ell$ \cite{cvetkovic2019near}. According to the spectrum in Eq. (\ref{eq7}), we obtain the following expression for the decay time of the modes associated with the test field in question
\begin{eqnarray}
\tau=\frac{e\mathcal{B}_0 \ell^2}{4mn^{*}}.
\end{eqnarray}
This result shows that decay time of the modes becomes longer as the strength of the external magnetic field increases. It is known that the decay time of such modes is related with the time to reach thermal equilibrium of the BHs \cite{konoplya2011quasinormal}. Hence, one can infer that the existence of such an external magnetic field in the near-horizon of the BHs may affect their thermalization process. Here, we should also underline that decay time for the modes corresponding to excited states of the considered perturbation test field is smaller than the decay time corresponding to ground state ($n=0$) mode provided that $m\neq 0$ (note that the sign of $\omega_{Im}$ remains unchanged when $m\longrightarrow-m$ and $e\longrightarrow-e$ and $\omega_{Im}=0$ when $m=0$). Our results have shown also that amplitude of the real oscillation modes decreases as the strength of the external magnetic field increases.

\section{Result and discussions}\label{sec:4}

In this contribution, we have analyzed the evolution of the test scalar field (with mass $\mu$ and charge $e$) exposed to an external uniform magnetic field in the near-horizon geometry of the BTZ BH with torsion. To acquire this, we have analytically solved the corresponding Klein-Gordon equation $\mathcal{KG}E$.  First of all, we have obtained the corresponding form of the $\mathcal{KG}E$ for the system in question. This equation is given in Eq. (\ref{eq4}). We have determined the solution function of this equation and accordingly have arrived at a frequency spectrum including real ($\omega_{Re}$) and imaginary ($\omega_{Im}$) parts in the form of $\omega=\omega_{Re}-i\omega_{Im}$ (see Eq. (\ref{eq7})). We observe that it does not seem to be possible to determine the characteristic oscillations of the considered geometric background under test scalar field perturbation in the absence of the external magnetic field. Our results mean also that the presence of an external uniform magnetic field can allow acquiring of information about the geometric background. In the Eq. (\ref{eq7}), the $\omega_{Re}$ gives the real oscillation frequency of the modes and the $\omega_{Im}$ (inverse) is proportional with their decay time. Our results have shown that the obtained frequency modes only oscillate if the azimuthal quantum number ($m$) of the particle is $m=0$ because of the fact that the imaginary part becomes $\omega_{Im}=0$ for this case. Provided that $m\neq0$, real oscillation frequency of the modes and their decay time depend on the strength of the external uniform magnetic field ($\propto \mathcal{B}_0$) besides the parameters of the geometric background. This fact allows us to analyze the effects of the external magnetic field on the obtained real oscillation modes and their decay times.

\section{Conclusions}\label{sec:5}

Here, we observe that the amplitude of the real oscillation modes decreases as the strength of the external magnetic field increases (see  Fig. \ref{fig1}). Our results show also that the decay time of the modes increases linearly as the strength of the external magnetic field increases (see Fig. \ref{fig3}). In our results, the $\omega_{Im}$ is positive and this means that the near-horizon geometry of the BTZ BH with torsion is stable under the considered perturbation field. One can also see that the decay time of the modes for the test field in question is independent of the value of the event horizon ($r_0$) even though the real oscillation frequency of the modes depends explicitly on the $r_0$. Our results imply that existence of the external magnetic field in the near-horizon of the BTZ BH with torsion may affect the time to reach a thermal equilibrium of the BH (see \cite{konoplya2011quasinormal}).

\section*{Acknowledgments}

The authors thank the anonymous reviewer for reading the manuscript, valuable comments and kind suggestions.

\section*{\small{Data availability}}
This manuscript has no associated data or the data will not be deposited.
\section*{\small{Conflicts of interest statement}}
There is no conflict of interest declared by the authors.
\section*{\small{Funding}}
There is no funding regarding this article.

\end{spacing}

\begin{thebibliography}{0}    

\bibitem{deser1984three}
S Deser,  R Jackiw,  G. 't Hooft  \textit{Ann. Phys.}, \textbf{152}, 220--235 (1984).

\bibitem{deser1988classical}
S Deser,  R Jackiw,  {\it Commun. Math. Phys.}, \textbf{118}, 495--509 (1988).

\bibitem{t1988non}
G. 't Hooft, {\it Commun. Math. Phys.}, \textbf{117}, 685--700 (1988).

\bibitem{witten19882+}
E Witten,  {\it Nuc. Phys. B}, \textbf{311}, 46--78 (1988).

\bibitem{witten1989topology}
E Witten ,{\it Nuc. Phys. B}, \textbf{323}, 113--140 (1989).

\bibitem{carlip19952+}
S Carlip, {\it Classical Quant. Grav.}, \textbf{12}, 2853 (1995).

\bibitem{barrow1986three}
 J D Barrow,  A B Burd, D Lancaster, {\it Classical Quant. Grav.}, \textbf{3}, 551 (1986).

\bibitem{banados1992black}
M Banados, C Teitelboim,  J Zanelli,{\it Phys. Rev. Lett.}, \textbf{69}, 1849 (1992).

\bibitem{konoplya2011quasinormal}
R A Konoplya, A Zhidenko, {\it Rev. Mod. Phys.}, \textbf{83}, 793 (2011).

\bibitem{cvetkovic2019near}
B Cvetkovi{\'c}, D Simi{\'c},{\it Phys. Rev. D}, \textbf{99}, 024032 (2019).

\bibitem{rincon2018scale}
 {\'A} Rinc{\'o}n, B Koch, {\it Eur. Phys. J. C}, \textbf{78}, 1--10 (2018).

\bibitem{gecim2018quantum}
G Gecim, Y Sucu, {\it Gen. Relativ. Gravit.}, \textbf{50}, 1--15 (2018).

\bibitem{xu2020diagnosis}
Z M Xu, B Wu, W L Yang, {\it Eur. Phys. J. C}, \textbf{80}, 1--10 (2020).

\bibitem{canate2020nonlinear}
P Ca{\~n}ate, D Magos, N Breton,{\it Phys. Rev. D}, \textbf{101}, 064010 (2020).

\bibitem{panotopoulos2017greybody}
G Panotopoulos, {\'A} Rinc{\'o}n, {\it Phys. Lett. B}, \textbf{772}, 523--528 (2017).

\bibitem{tekincay2021exotic}
C Tekincay, M Dernek, Y Sucu, {\it Eur. Phys. J. Plus}, \textbf{136}, 1--13 (2021).

\bibitem{panotopoulos2018quasinormal}
G Panotopoulos, {\it Gen. Relat. Gravit.}, \textbf{50}, 1--13 (2018).

\bibitem{suggestion} R Becar, P A Gonzalez, Y Vasquez, {\it Phys. Rev. D} \textbf{89}, 023001 (2014).

\bibitem{fathi2021adiabatic}
M Fathi, S Lepe, J R Villanueva, {\it Eur. Phys. J. C}, \textbf{81},1--9 (2021).

\bibitem{sfetsos1998microscopic}
K Sfetsos, K Skenderis, {\it Nuc. Phys. B}, \textbf{517}, 179--204 (1998).

\bibitem{strominger1998black}
A Strominger, {\it J. High Energy Phys.}, \textbf{1998}, 009 (1998).

\bibitem{birmingham2001exact}
D Birmingham, I Sachs, S Sen, {\it Int. J. Mod. Phys. D}, \textbf{10}, 833--857 (2001).

\bibitem{hawking1974black}
S W Hawking, {\it Nature}, \textbf{248}, 30--31 (1974).

\bibitem{hawking1975particle}
S W Hawking,  \textit{Commun.Math. Phys.} 43, 199–220 (1975).

\bibitem{regge1957stability}
T Regge, J A Wheeler, {\it Phys. Rev.}, \textbf{108}, 1063 (1957).

\bibitem{zerilli1970effective}
F J Zerilli, {\it Phys. Rev. Lett.}, \textbf{24}, 737 (1970).

\bibitem{zerilli1974perturbation}
F J Zerilli,  {\it Phys. Rev. D}, \textbf{9}, 860 (1974).

\bibitem{Chandrasekhar}
S Chandrasekhar, \textit{The Mathematical Theory of Black Holes.} \textit{Clarendon, Oxford University Press}, R. H. Romer, 646 pp., (1985)

\bibitem{berti2009quasinormal}
E Berti, V Cardoso, A O Starinets,  \textit{Class. Quant. Grav.}, \textbf{26}, 163001 (2009).

\bibitem{Vieira}
H S, Vieira, K D Kokkotas,  {\it Phys. Rev. D}, \textbf{104}, 024035 (2021).

\bibitem{Yang}
Y Huang, H Zhang, {\it Phys. Rev. D}, \textbf{103}, 044062 (2021).

\bibitem{Hod}
S Hod, {\it Phys. Lett. B}, \textbf{749}, 167-171 (2015)

\bibitem{Marco}
M O P Sampaio, C Herdeiro, M Wang, {\it Phys. Rev. D} \textbf{90}, 064004 (2014).

\bibitem{guica2009kerr}
M Guica, T Hartman, W Song, A Strominger,{\it Phys. Rev. D}, \textbf{80}, 124008 (2009).

\bibitem{cunha2016relativistic}
M S Cunha, C R Muniz, H R Christiansen, V B Bezerra, {\it Eur. Phys. J. C}, \textbf{76}, 1--7 (2016).

\bibitem{vitoria2018rotating}
V B Vit{\'o}ria, K Bakke, {\it Eur. Phys. J. C}, \textbf{78}, 1--6 (2018).

\bibitem{zare2020non}
S Zare, H Hassanabadi, M de Montigny, {\it Gen. Relativ. Gravit.}, \textbf{52}, 1--20 (2020).

\bibitem{guvendi2021relativisticb}
A Guvendi, H Hassanabadi,{\it Few-Body Sys.}, \textbf{62}, 1--8 (2021).

\bibitem{guvendi2021effects}
A Guvendi,  {\it Sakarya Uni. J. Sci.}, \textbf{25}, 847--853 (2021)

\bibitem{shie2008torsion}
Kun-F Shie, J M Nester, Hwei-J  Yo, {\it Phys. Rev. D}, \textbf{78}, 023522 (2008).

\bibitem{belyaev2017torsion}
A S Belyaev, M C Thomas, I L Shapiro,{\it Phys. Rev. D}, \textbf{95}, 095033 (2017).

\bibitem{tilquin2011torsion}
A Tilquin, T Sch{\"u}cker, {\it Gen. Relativ. Gravit.}, \textbf{43}, 2965 (2011).

\bibitem{minkevich2009accelerating}
A V Minkevich, {\it Phys. Lett. B}, \textbf{678}, 423--426 (2009).

\bibitem{enqvist1994ferromagnetic}
K Enqvist, P Olesen, {\it Phys. Lett. B}, \textbf{329}, 195--198 (1994).

\bibitem{grasso2001magnetic}
D Grasso, H R Rubinstein, {\it Phys. Rep.}, \textbf{348}, 163--266 (2001).

\bibitem{guvendi2021relativisticC}
A Guvendi, S Gurtas Dogan, {\it Few-Body Sys.}, \textbf{62}, 1--8 (2021).

\bibitem{guvendi2021relativistic}
A Guvendi, {\it Eur. Phys. J. C}, \textbf{81}, 1--7 (2021).

\bibitem{dietl2008new}
P Dietl, F Pi{\'e}chon, G Montambaux,  {\it Phys. Rev. Lett.}, \textbf{100}, 236405 (2008).

\bibitem{furtado1994landau}
C Furtado, B G C da Cunha, F  Moraes, E R B de Mello,  V B Bezzerra, {\it Phys. Lett. A}, \textbf{195}, 90--94 (1994).

\bibitem{medeiros2012relativistic}
E R F Medeiros, E R B de Mello, {\it Eur. Phys. J. C}, \textbf{72}, 1--14 (2012).

\bibitem{hammad2020landau}
F Hammad, A Landry, {\it Eur. Phys. J. Plus}, \textbf{135}, 1--23 (2020).

\bibitem{sakalli2004solution}
I Sakalli, M Halilsoy,  {\it Phys. Rev. D}, \textbf{69}, 124012 (2004).

\bibitem{hassanabadi2017duffin}
H Hassanabadi, M  Hosseinpour, M De Montigny, {\it Eur. Phys. J. Plus}, \textbf{132}, 1--12 (2017).

\bibitem{guvendi2021vector}
A Guvendi, S Zare, H Hassanabadi, {\it Eur. Phys. J. A}, \textbf{57}, 1--6 (2021).

\bibitem{guvendi2020interacting}
A Guvendi, Y Sucu,  {\it Phys. Lett. B}, \textbf{811}, 135960 (2020).

\bibitem{ahmed2020generalized}
F Ahmed, {\it Eur. Phys. J. C}, \textbf{80}, 1--12 (2020).

\bibitem{dogan2021two}
S G Dogan, {\it Sakarya Uni. J. Sci.}, \textbf{25}, 1210--1217 (2021)

\bibitem{guvendi2021Dynamics}
A Guvendi, {\it Int. J. Mod. Phys. A}, \textbf{36}, 2150144 (2021).

\bibitem{guvendinoninertial}
A Guvendi, H Hassanabadi, {\it Int. J. Mod. Phys. A}, \textbf{36}, 2150253 (2021)

\bibitem{santos2018relativistic}
L C N Santos, C C Barros,  {\it Eur. Phys. J. C}, \textbf{78}, 1--8 (2018).

\bibitem{cardoso2001scalar}
V Cardoso, J P S Lemos, {\it Phys. Rev. D}, \textbf{63}, 124015 (2001).

\bibitem{horowitz2000quasinormal}
G T Horowitz, V E Hubeny,  {\it Phys. Rev. D}, \textbf{62}, 024027 (2000).

\bibitem{konoplya2006stability}
R A Konoplya, A V Zhidenko,  {\it Phys. Rev. D}, \textbf{73}, 124040 (2006).

\bibitem{konoplya2002decay}
R A Konoplya, {\it Phys. Rev. D}, \textbf{66}, 084007 (2002).


\bibitem{detweiler1980klein}
S Detweiler,  {\it Phys. Rev. D}, \textbf{22}, 2323 (1980).


\bibitem{KGO}
A R Soares,  R L L Vit{\'o}ria, H Aounallah,  {\it Eur. Phys. J. Plus}, \textbf{136},  1--8 (2021).

\bibitem{zare2022interaction}
S Zare, H Hassanabadi, A Guvendi, W S Chung,  {\it Int. J. Mod. Phys. A}, \textbf{37}, 2250033 (2022) .

\bibitem{turimov2019quasinormal}
B Turimov, B Toshmatov, B Ahmedov, Z Stuchl{\'\i}k, {\it Phys. Rev. D}, \textbf{100}, 084038  (2019).

\bibitem{konoplya2008quasinormal}
R A Konoplya, R D B Fontana, {\it Phys. Lett. B}, \textbf{659}, 375--379   (2008).

\bibitem{wu2015decay}
C Wu, R Xu,  {\it Eur. Phys. J. C}, \textbf{75},  1--5  (2015).

\bibitem{kokkotas2011quasinormal}
K D Kokkotas, R A Konoplya, A Zhidenko,{\it Phys. Rev. D}, \textbf{83}, 024031   (2011).

\bibitem{dogan2019quasinormal}
S G Dogan, Y Sucu, {\it Phys. Lett. B}, \textbf{797}, 134839   (2019).

\bibitem{book1}
G B Arfken, H J Weber, F E Harris, {\it Mathematical Methods for Physicists, Seventh Edition: A Comprehensive Guide}, \textit{ Cambridge, Academic Press}, 1206,  (2012)

\bibitem{torsionnew}
R L L Vit{\'o}ria, K Bakke,  {\it Gen. Relativ. Gravit.}, \textbf{48}, 1--11 (2016).
\end{thebibliography}
\end{document}